\documentclass[10pt,conference]{IEEEtran}
\ifCLASSINFOpdf
  \usepackage[pdftex]{graphicx}
\else
\fi
\usepackage{amsmath}

\usepackage{cleveref}
\usepackage{xcolor}

\begin{document}
%
\title{A Scalable FPGA Architecture for Real-Time Decoding of Quantum LDPC Codes Using GARI}
%
%
%
\author{Daniel Báscones$^{(1)}$,
        Arshpreet Singh Maan$^{(2)}$, Valentin Savin$^{(3)}$
        and~Francisco Garcia-Herrero$^{(1)}$\\[3mm]
        $^{(1)}$\,Complutense University of Madrid, Madrid, Spain, \{danibasc, francg18\}@ucm.es\\
        $^{(2)}$\,Aalto University, Espoo, Finland, arshpreet.maan@aalto.fi\\
        $^{(3)}$\,Quobly, Grenoble, France, valentin.savin@quobly.io
}

%
%

\markboth{IEEE International Conference on Quantum Computing \& Engineering (QCE 2026)}%
{Shell \MakeLowercase{\textit{et al.}}: Bare Demo of IEEEtran.cls for IEEE Journals}
%



\maketitle

\begin{abstract}
In this work, we introduce a new hardware architecture for decoding correlated errors in quantum LDPC codes. The decoder is based on message passing and exploits the structure of the detector error model obtained through the recently introduced Graph Augmentation and Rewiring for Inference (GARI) method. The proposed architecture enables flexible scaling and can, in principle, adapt to any quantum LDPC codes using the GARI framework. It leverages resource reuse while maintaining a modest degree of parallelism, thereby reducing power consumption and area requirements, while preserving low decoding latency.  

As a case study, 
the architecture was implemented on a VCU19P FPGA as an ensemble of three decoder cores targeting the [[144,12,12]] bivariate bicycle code,  achieving an average latency of 596\,ns per decoding round. This implementation consumes six times fewer resources than the previous GARI-based proposal, being the first reported implementation of multiple decoder cores for correlated errors on a single FPGA device.
This enables better energy-conscious scaling of the quantum error correction layer on the classical side, reducing overall power consumption while meeting real-time constraints without compromising decoding accuracy under correlated errors.

\end{abstract}


%
\IEEEpeerreviewmaketitle

\section{Introduction}

The realization of real-time decoders for quantum error correction (QEC) remains a challenge on the path to large-scale fault-tolerant quantum computing \cite{mohseni2025buildquantumsupercomputerscaling}, \cite{palsberg2026computersciencechallengesquantum}. This challenge becomes particularly acute when scalability in real-time decoding is pursued by increasing the maximum code distance while simultaneously adopting more realistic and complete error models, including correlated errors \cite{beni2025tesseractsearchbaseddecoderquantum}. In this regime, decoding performance is no longer determined only by algorithmic accuracy, but also by latency, classical computing resources, and power constraints imposed by the underlying hardware platform  \cite{mohseni2025buildquantumsupercomputerscaling}, \cite{fellous2023optimizing}.

Surface codes were the first family to confront these constraints at scale, and consequently, a significant body of hardware-oriented research has emerged around them, such as \cite{ravi2022betterworstcasedecodingquantum}, \cite{liyanage2023scalablequantumerrorcorrection}, \cite{vittal2023astrea}, \cite{Barber_2025}, and \cite{Wu_2025}. In contrast, other quantum LDPC codes, such as bivariate bicycle codes \cite{Bravyi_2024}, have comparatively few reported real-time hardware implementations to date \cite{maurer2025realtimedecodinggrosscode}, \cite{bascones2025exploring}, some of which rely on estimations rather than full implementations \cite{maurya2026fpgatailoredalgorithmsrealtimedecoding}. 

Among the various proposed surface code decoder architectures, Micro Blossom stands out as one of the most prominent hardware-oriented approaches, as it obtains high accuracy for uncorrelated errors, while maintaining the maximum frequency when the distance scales \cite{Wu_2025}. For surface codes, Micro Blossom has been proposed to achieve average latencies per syndrome round below $1\mu$s on modern FPGA platforms such as the Versal VMK180 \cite{amd_vmk180}. These solutions typically exploit heterogeneous architectures that combine embedded CPUs and programmable logic (PL), interconnected through AXI buses. 
To date, implementations on a single FPGA have been demonstrated only for distances up to $d=13$, highlighting the challenges of scaling beyond this range.

On the other hand, implementations targeting bivariate bicycle codes have largely focused on FPGA realizations of belief propagation (BP)–based algorithms \cite{maurer2025realtimedecodinggrosscode} and some associated post-processing techniques such as ordered statistics decoding (OSD) \cite{bascones2025exploring}. The analyses in~\cite{maurya2026fpgatailoredalgorithmsrealtimedecoding} suggest that BP-based approaches are the most promising candidates for real-time decoding, compared to OSD and clustering-based alternatives such as~\cite{delfosse2022towardUF}, which exhibit longer latency tails in hardware implementations. This remains the case even for new OSD approximations such as filtered-OSD~\cite{maurya2026fpgatailoredalgorithmsrealtimedecoding}, where the latency, measured in clock cycles, can be comparable to that of three iterations of serial BP decoding.

In addition, despite the previous advances, both BP and OSD architectures encounter a fundamental scalability bottleneck for bivariate bicycle codes, as reported in \cite{maurer2025realtimedecodinggrosscode}, \cite{bascones2025exploring}, \cite{maurya2026fpgatailoredalgorithmsrealtimedecoding}. On some of the largest currently available FPGA evaluation platforms \cite{amd_vu19p_fpga}, maximum parallelism is effectively reached at distance $d = 12$, where the implementation of just one BP decoder with parallel schedule nearly exhausts available device resources \cite{maurer2025realtimedecodinggrosscode}. In this regime, constraints on Block RAM (BRAM) availability, routing congestion, and interconnect complexity limit the feasibility of fully parallel decoding of a complete detector error model graph that considers correlated errors with $d$ rounds. Even when sufficient logical resources remain, increased routing pressure and memory resources often cause a significant degradation in achievable operating frequency. As a result, simply scaling parallelism with code distance does not provide a sustainable path forward. Thus, even though bivariate bicycle codes offer higher efficiency than surface codes for memory experiments, both are limited to similar distances when decoding on a single FPGA device with sub-$1\mu$s latency. 

An additional and often underexplored limitation concerns numerical precision. Existing studies typically provide only limited analyses of quantization effects, and the impact of reduced precision on full detector error model graphs with correlated errors remains largely unknown for low logical error rates \cite{maurer2025realtimedecodinggrosscode}. Without a systematic characterization of these effects, it is not possible to determine the minimum bit width required to prevent error floors or to reliably mitigate them \cite{garciaherrero2025diversitymethodsimprovingconvergence}. This uncertainty further complicates architectural design decisions, as precision directly influences memory usage, routing pressure, and overall resource consumption.

Multi-FPGA solutions have been proposed as a potential strategy to overcome single-device resource limitations, particularly for ensemble-based BP decoder architectures \cite{koutsioumpas2025automorphismensembledecodingquantum}, \cite{koutsioumpas2025colourcodesreachsurface}, \cite{maan2025decodingcorrelatederrorsquantum}. While conceptually attractive, such direct scaling introduces additional challenges. Beyond inter-device communication complexity, power consumption becomes a critical concern \cite{meier2025energy}. As highlighted in~\cite{Auff_ves_2022}, the classical infrastructure required to support large-scale fault-tolerant quantum processors, including control electronics and real-time decoding layers, will constitute a non-negligible fraction of the total system power budget. When architectures must be replicated multiple times to manage distinct groups of logical qubits, power efficiency becomes a first-order design constraint rather than a secondary consideration.

Motivated by these limitations, this work proposes an architecture designed to meet strict real-time latency requirements while correcting correlated errors using the graph augmentation and rewiring for inference (GARI) technique \cite{maan2025decodingcorrelatederrorsquantum}. Instead of pursuing fully parallel multi-FPGA implementations \cite{valls2021syndromebased}, the proposed approach emphasizes controlled parallelism and systematic resource reuse. By carefully balancing latency, accuracy, and hardware utilization, the architecture aims to reduce the number of required devices in a potential decoder ensemble while simultaneously lowering overall power consumption. In doing so, it provides a scalable and energy-conscious pathway toward real-time decoding of correlated error models in quantum LDPC codes.

The remainder of this paper is organized as follows. Section II provides background on the GARI technique introduced in \cite{maan2025decodingcorrelatederrorsquantum}. Section III describes the proposed architecture in detail. Section IV presents the resource utilization and timing results for a single decoder, as well as for an ensemble that fits within a single FPGA for the [[144,12,12]] bivariate bicycle code
 code. Finally, Section V summarizes the conclusions and future work.

\section{Background}

Quantum LDPC codes are a promising class of error-correcting codes for fault-tolerant quantum computing \cite{Panteleev_2021, Bravyi_2024}. However, their decoding is particularly challenging due to the presence of both $X$ and $Z$ error components, as well as correlations between these error types arising from $Y$ errors. Conventional decoding approaches often treat $X$ and $Z$ errors independently, which leads to suboptimal performance in the presence of correlated noise \cite{beni2025tesseractsearchbaseddecoderquantum}.

The GARI method proposed in \cite{maan2025decodingcorrelatederrorsquantum} modifies the correlated detector error model by eliminating 4-cycles involving $Y$-type errors. Starting from the original detector error model, which can be written as:

\begin{equation}\label{eq:original_matrix}
\overline{D}_{XYZ} =
\begin{matrix}
\begin{matrix}
e_Z & e_X & e_Y 
\end{matrix} & \\
\begin{pmatrix}
D_X & 0 & D'_X \\
0 & D_Z & D'_Z \\
\end{pmatrix} &
\begin{matrix}
s_X \\
s_Z \\
\end{matrix} \\
\end{matrix}
\end{equation}
the columns represent the error nodes (also called variable nodes), and the rows represent the detector nodes (also called syndromes or check nodes). It is assumed that $D_X$ and $D_Z$ have no repeating columns, since errors producing the same syndrome are merged. 
Furthermore, each column of $D_X'$ corresponds to a column of $D_X$, and each column of $D_Z'$ corresponds to a column of $D_Z$, though repetitions may occur. These repetitions capture the effect of $Y$ errors occurring at different locations that produce the same $X$-syndrome but distinct $Z$-syndromes, and vice versa.

These relationships imply the existence of matrices $U$ and $V$ such that
\[
D_X' = D_X U, \qquad D_Z' = D_Z V,
\]
where $U$ and $V$ encode the repetitions of columns in $D_X'$ and $D_Z'$, respectively. Using this structure, one can perform a change of variables so that the detector error model can be rewritten as in \Cref{eq:GARI_matrix}.

  \begin{equation}\label{eq:GARI_matrix}
    \overline{D}_{XYZ} =
        \begin{matrix}
        \begin{matrix}
        e_Z & e_X \ & e_Y \ & \overline{e}_Z & \overline{e}_X
        \end{matrix} & \\
        \begin{pmatrix}
        \ 0 \ & \ 0 \ & \ 0 \ & D_X & 0 \\
        \ 0 \ & \ 0 \ & \ 0 \ & 0 & D_Z \\
        \ I \ & \ 0 \ & \ U \ & I & 0 \\
        \ 0 \ & \ I \ & \ V \ & 0 & I
        \end{pmatrix} &
        \begin{matrix}
        s_X \\
        s_Z \\
        0 \\
        0
        \end{matrix} \\
        \end{matrix}
    \end{equation}

Conceptually, the GARI transform breaks short cycles by separating correlated components and restructuring the decoding problem through $U$ and $V$, while enabling joint decoding through the exchange of reliability information between the two error domains.
In this way, harmful correlations between messages are mitigated, improving the convergence and reliability of iterative decoding, as demonstrated in~\cite{maan2025decodingcorrelatederrorsquantum}.

\section{Architecture proposal}

    \subsection{Overview} 

    The GARI architecture is based on two different decoders, which derive from the inherent properties of the $\bar{D}_{XYZ}$ matrix \cite{maan2025decodingcorrelatederrorsquantum} shown in \Cref{eq:GARI_matrix}. As with any code, the ordering in which the different checks -or rows- are processed can greatly influence the decoding result \cite{ducrest2023layered}. For our particular case, we notice that there are dependencies between $D_X$, $D_Z$, $U$, and $V$ through the different variables and checks. $\bar{e}_Z$ are shared between $D_X$ and $U$, $\bar{e}_X$ between $D_Z$ and $V$, and $e_Y$ between $U$ and $V$. This will later prompt separating checks into the different processing units.

\begin{figure}
        \centering
        \includegraphics[width=0.75\linewidth]{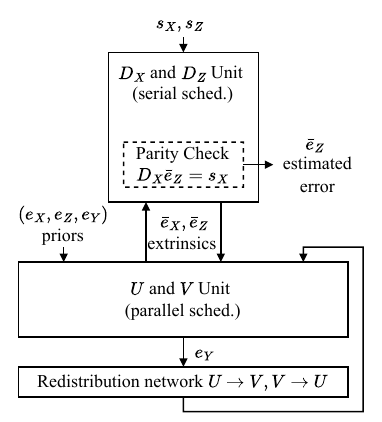}
        \caption{Overview of the GARI architecture from this work.}
        \label{fig:GARI_arch_overview}
    \end{figure}

    Note that $D_X$ and $D_Z$ have a fairly similar structure (intertwined checks with a degree of tens of variables), while $U$ and $V$ have fully independent checks of fewer variables. A serial BP schedule fits the first pair nicely, while a more parallel structure can be devised for the second due to the independent nature of the checks. Furthermore, $D_X,V$ and $D_Z,U$ can be processed simultaneously as no direct information exchange takes place between the pairs. These facts prompt the structure shown in \Cref{fig:GARI_arch_overview}.

    A serially scheduled BP decoder processes $D_X$ and $D_Z$ (henceforth \emph{$D_X,D_Z$ unit}). Using ping-pong-like buffers, only one is processed at a time while outputting data to the \emph{$U,V$ unit}. The same repeats for the \emph{$U,V$ unit}, where only one matrix is active in processing. Data dependencies between these submatrices are solved using a network of interconnections between the memories storing their respective value nodes. The following expressions show the data flow that takes place between these alternating cycles:

    \begin{equation}
        D_X\to U \quad V \to U, D_Z
    \end{equation}
    \begin{equation}
        D_Z\to V \quad U \to V, D_X
    \end{equation}

    Assuming the decoding process starts with $D_X$ (starting with $D_Z$ would be symmetric), then we have the timeline for processing shown in \Cref{tab:opscheduling}. Note that the parity checks are carried out only over $D_Z$ (for memory experiments), so the decoder can only converge in an even number of steps.

    \begin{table}[h]
        \centering
        \caption{Scheduling of the decoding operations}
        \label{tab:opscheduling}
        \begin{tabular}{c|c|c|c|c|c|c}
        Step & 1 & 1' & 2 & 2' & 3 & \dots  \\\hline\hline
        Serial decoder      & $D_X$ & $D_Z$ & $D_X$ & $D_Z$ & $D_X$ & \dots\\
        Parallel decoder    &       &  $U$  &  $V$  &  $U$  &  $V$  &  \dots \\
        \end{tabular}
    \end{table}

    \subsection{The \emph{$D_X,D_Z$ unit}} 

        \begin{figure*}
            \centering
            \includegraphics[width=\linewidth]{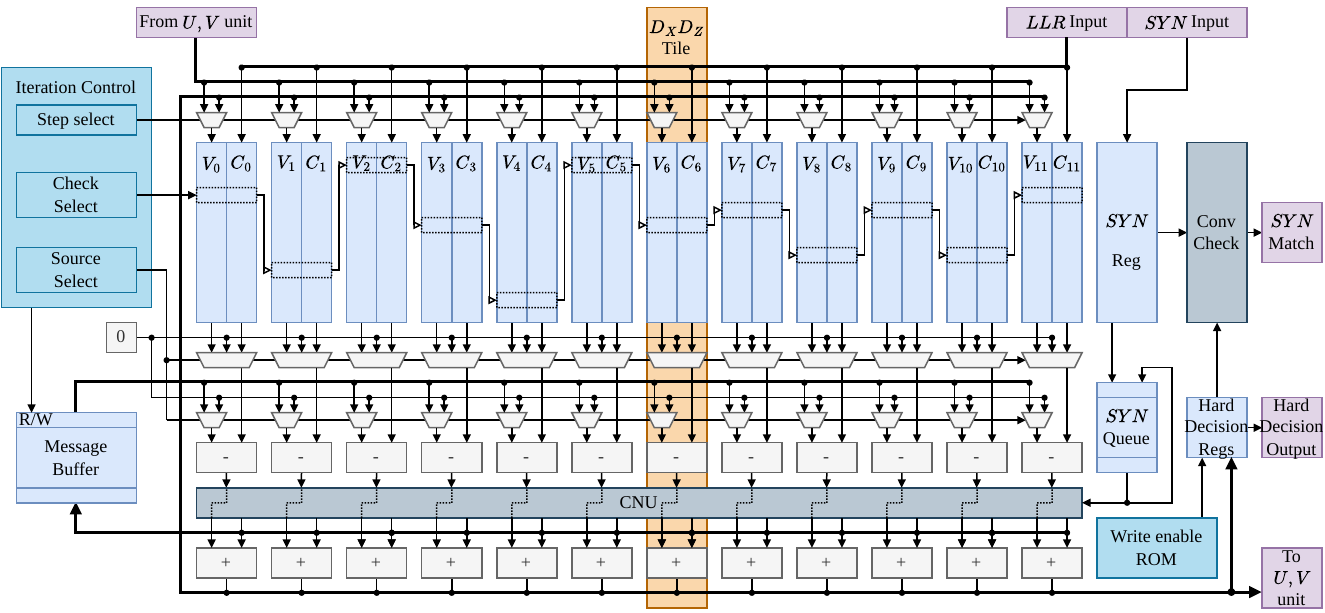}
            \caption{The architecture for the $D_X,D_Z$ decoder. Memory elements shown in blue (light for RAM, dark for ROM). I/O in purple. Sub-modules in gray.}
            \label{fig:dxdzdecoder}
        \end{figure*}
    
        BP decoders are often used for their speed and resource efficiency. Two options are often considered: parallel and serial scheduled decoders. 
        The former~\cite{valls2021syndromebased} performs fully parallel updates in an alternating schedule, where all variable nodes updated simultaneously followed by all check nodes.
        While extremely fast in theory, it has three main drawbacks: 1) the high number of required registers and interconnects makes it difficult to implement and scale for codes with larger distances or detector error model graphs with higher complexity \cite{maurer2025realtimedecodinggrosscode}, \cite{koutsioumpas2025colourcodesreachsurface}, 2) the decoding performance is subpar when compared to serial scheduled decoders \cite{ducrest2023layered} \cite{maan2025decodingcorrelatederrorsquantum} and 3) the code is embedded into the logic, which for our case would mean two different decoders would be necessary.
    
        Thus, a serial decoder is chosen for the purpose of decoding $D_X$ and $D_Z$, the \emph{$D_X,D_Z$ unit}. Register-based and memory-based \cite{roth2011area} serial BP decoders are the main alternatives, which differ in how the variable information is stored. For the former, each variable node is stored in a different register. This requires a huge array of interconnects and multiplexers to route all possible combinations, corresponding to the different checks, from the variable registers towards the update units. This, again, is not scalable and fixes the code in the logic itself. 
        The alternative is memory-based decoders, in which each variable is assigned a memory. Memories are shared under the constraint that variables connected to the same checks must reside in different memories.  Given the large number of variables in typical codes, this results in an NP-hard optimization problem that is not trivial to solve. 
        Fortunately, as we will show, a high-performing decoder can be designed without requiring an optimal assignment of variables to memories.
    
    
        The full proposed architecture for the \emph{$D_X,D_Z$ unit} is shown in \Cref{fig:dxdzdecoder}. For simplicity, the figure shows a maximum check-node degree of 12. 
        Note the repeating structure called \emph{$D_X,D_Z$ tile}, which contains both the memory and logic for the update of a variable subset. The unit can trivially scale up if needed by adding more tiles.
        
        \subsubsection{Memory structure}
            Two memory arrays are used for the LLR values. Value memories $V_i$ can read LLR values for processing one check while writing updated LLR values resulting from a previously processed check, delayed by the internal pipeline depth. 
            At startup, initial LLR values obtained from device calibration must be injected, which could, in principle, be handled within the same memories using input multiplexing. However, this would require resetting the LLRs at each run, which is cycle-inefficient. To avoid this, we use a separate array of calibration memories $C_i$, enabling on-the-fly calibration to adapt the decoder to drift~\cite{Proctor_2020}, while the value memories $V_i$ are reserved for the decoding process. 

            Check-node messages are stored in a FIFO buffer and used in the next iteration. 

            Syndromes are stored both in a register file for parity checking and in a circular queue for check processing.
    
        \subsubsection{CNU processing}
            As the degree of the nodes in the graph of the detector error model is usually irregular, the inputs and outputs of the CNU (check-node unit) are not always fully utilized. This necessitates masking unused values when a check-node degree is smaller than the maximum, so the CNU inputs include dedicated filters for this purpose. Output filtering is not required, since downstream processing is protected by the input stage, which blocks invalid values. 
            Within the CNU, message updates are performed according to the normalized min-sum rules~\cite{savin2014ldpc}, exchanging information between all connected variables.
    
        \subsubsection{Control unit}
            To control the hardware, an ``iteration control'' unit is devised. Depending on the iteration and scheduling step, it reads an array of ROMs to select:

            \begin{itemize}
                \item The input source for the variable nodes: They can either be updated from a local iteration within the $D_X$ and $D_Z$ matrices or from the update coming from $U$ and $V$. To enable alternating processing of $D_X$ and $D_Z$ on the same hardware, the memories are duplicated $V_i^{D_X}, V_i^{D_Y}$ so that when one is used for processing, the other is feeding from the $U,V$ unit.
                \item The variable nodes used for the current check: The check is formed by selecting the participating variables via a set of indices, one for each memory, coming from the control ROM. 
                \item The source for the CNU: Depending on the iteration, the CNU source for each variable can be either $V_i$ or $C_i$, which in any case are combined with the buffered message from the previous iteration. Any values that are unused for that specific check are masked out and replaced with maximums so that they do not interfere in the min-sum update. These are all controlled with another ROM that stores the select bits for the mux depending on the iteration.
                \item The enable signal for the hard decision registers: Only a small number of variables are updated on each scheduling step, so a ROM containing enable signals is used to select them.
            \end{itemize}

        \subsubsection{Convergence checking unit}
            Convergence checking is performed in two steps: First, information from the hard decision registers and syndrome registers is combined according to the parity matrix graph, producing the parity results for each check. In a second pipelined step, these individual parities are aggregated to to determine whether any violations remain. If all checks are satisfied, decoding is terminated and the hard decision is output.

            For the case of the $[[144,12,12]]$ code, and since we only need to compute the parity checks corresponding to either $D_X$ or $D_Z$ (depending on the considered memory experiment), this approach is expected to be sufficient. Thus, for the results in \Cref{sec:restiming}, we use this fully parallel approach.
            
            
    \begin{figure*}
        \centering
        \includegraphics[width=\linewidth]{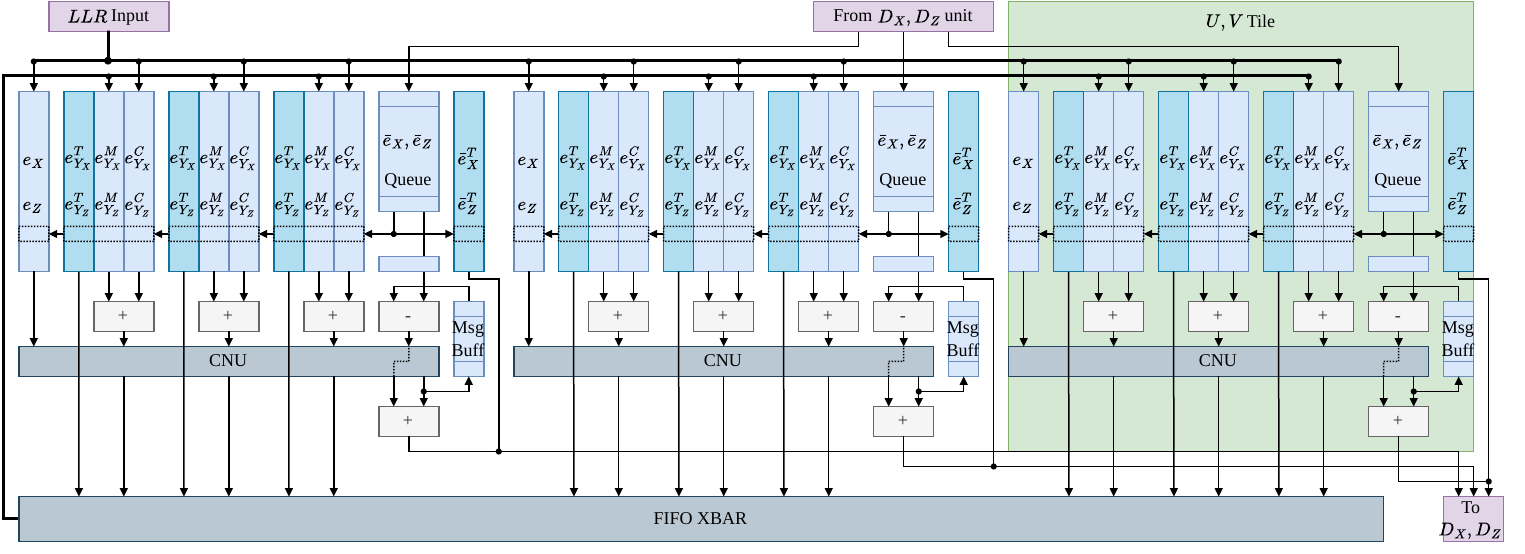}  
        \caption{Architecture for the $U,V$ decoder. Memory elements shown in blue (light for RAM, dark for ROM), I/O in purple. Sub-modules in gray.}
        \label{fig:uvdecoder}
    \end{figure*}

    \subsection{The \emph{$U,V$ unit}} 

        The \emph{$U,V$ unit} exploits an aspect of the internal structure of this part of the decoding matrix: all checks are independent within $U$ and $V$, with dependencies existing between, but not within, the two matrices. This allows full parallelization when processing each matrix individually. Given the processing time for the \emph{$D_X,D_Z$ unit}, the degree of parallelization is chosen such that the processing time of $U$ and $V$ matches that of $D_X$ and $D_Z$. This avoids either idle time or unnecessary resource usage (hence power consumption) that would result from under- or overestimating the degree of parallelization.
    

Considering the structure of the GARI matrix $\overline{D}_{XYZ}$ in Equation~\eqref{eq:GARI_matrix}, any check associated with the lower block rows involving $U$ and $V$ receives messages from three types of variables: 
\begin{itemize}
    \item A variable from either $e_Z$ or $e_X$, whose message remains fixed at its initial LLR value.

    \item Variables from $e_Y$, each connected to exactly two checks (one in $U$ and one in $V$). This property can be exploited to simplify the hardware, since their messages are equal to the initial LLR value plus the incoming message from the opposite matrix update.

    \item A single variable node from either \(\bar{e}_X\) or \(\bar{e}_Z\), whose message must be brought in from the \emph{$D_X, D_Z$ unit}.
\end{itemize}
Note that any processing order within $U$ or $V$ can be used and yields the same computed messages (since checks within each matrix are independent).
    
        The input LLR values 
        stored in calibration memories of the $U,V$ unit are denoted by $e^C_i$ to distinguish them from those in the $D_X$ and $D_Z$ units.
        In contrast to the input LLR memories from the $D_X$ and $D_Z$ units, these cannot be updated online since their base values are used in all iterations. 
        This is not a problem since, in the startup sequence, the first iteration only processes $D_X$, and the loading of $U$ and $V$ LLRs can be overlapped. 
    
        The CNU messages of the $U,V$ unit are stored in message memories named as $e^M_i$. 
        These flow from $U$ to $V$ through a crossbar-like structure \cite{kanizo2009crosspoint} is essentially an $N$-to-$N$ router. 
        In principle, mapping $U$ outputs to $V$ inputs (and vice-versa) could be achieved by carefully assigning checks to check-node processors, thereby simplifying the intermediate routing logic. However, in practice, the resulting relationships are too complex for this to be efficient. Instead, we assign each outgoing message a tag indicating its destination memory $e_i^M$, generated by the tag memories $e^T_i$. 
        This information is processed by the \emph{crossbar} module in order to properly sort and distribute the messages. 
        
        The benefit of this approach is a controller-less structure that operates purely on data availability in the input queues, abstracting away explicit control logic. A minor overhead arises 
        when multiple outputs map to the same input during certain iterations, requiring additional cycles to drain the buffers. However, this is not a practical limitation, as confirmed by simulations. Further details are provided in \Cref{sec:restiming}.

        \subsubsection{The Crossbar interconnect}

            Routing $e_Y$ messages from $U$ to $V$ is done via the crossbar interconnect. For the $[[144,12,12]]$ gross code, there are already thousands of checks on each $U$ and $V$, each containing a different subset of $e_Y$. For our case, we assume no patterns can be extracted for optimizing routing, and that any \emph{$U,V$ tile} output can map to any \emph{$U,V$ tile} input through the interconnect. Note that some codes may exhibit patterns that avoid a $J$-to-$J$ routing, with $J$ being the number of I/Os across all \emph{$U,V$ tiles}.

            This type of routing is quite expensive to perform. First, having enough multiplexers is a challenge in itself; additionally, due to the mapping of checks, it may be the case that it is not $J$-to-$J$ in all stages but $J$-to-$(J-i)$ for some small $i$. Since latency is not critical at this stage, as long as processing completes before the \emph{$D_X,D_Z$ unit}, this motivates a layered structure that iteratively sorts messages according to their destination, given by a tag: $e^T_{Y_X}, e^T_{Y_Z}$. The first tag refers to messages exchanged between the $e_Y$ and $\bar{e}_X$ variable nodes, while the second tag refers to messages exchanged between the $e_Y$ and $\bar{e}_Z$ variable nodes. 
            Since the tag values are bounded by the number of inputs to the $U,V$ processors, 
            routing becomes similar to binary radix sorting~\cite{cormen2022introduction} an array in stages.

            First, to avoid stalling the \emph{$U,V$ tiles} outputs, the outgoing $J$ data pairs (message, tag) are received, after the first crossbar level, by FIFOs that are sized to absorb 
            any back-pressure. Interconnecting is done in $K=\lceil\log_2(J)\rceil$ stages. Each stage consists of an input crossbar, a $J/2$ distribution module array (each feed by two inputs 
            ), and an output crossbar feeding two outputs. These outputs are stored in ping-pong buffers in the intermediate stages, with an additional set of FIFOs at the final output. 
            The distribution modules input indices are separated by $2^{K-1-i}$, where $i=0\dots K-1$ is the current stage. 
            Each module routes messages based on the most significant bit (MSB) of the input tag, directing them either to the left or right output.
            At stage $i$, the crossbar connections for index $j$ follow the mapping in \Cref{eq:xbar}.
            \begin{equation}\label{eq:xbar}
                j\to (j\bmod 2^{K-1-i})\cdot2+\left\lfloor\frac{J}{2^{K-1-i}}\right\rfloor\bmod 2 + \left\lfloor\frac{J}{2^{K-i}}\right\rfloor\cdot 2^{K-i}
            \end{equation}

\begin{figure}
                \centering
                \includegraphics[width=\linewidth]{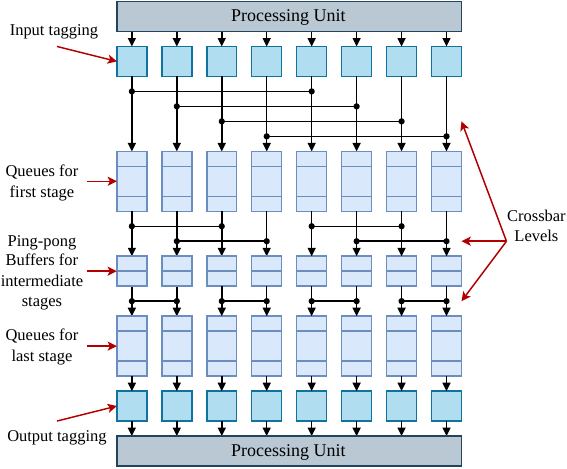}
                \caption{The crossbar interconnect. Note that the inputs and outputs are buffered in FIFO queues to relieve pressure from the input and output modules. Internally, ping-pong buffers suffice.}
                \label{fig:fifoxbar}
            \end{figure}

            At the output, messages are written to a memory module for storage. 
            The target memory index is read from a ROM, since the message ordering is fully deterministic. 
            An alternative would be to embed the target index in the tag prior to the crossbar stage. However, this would increase internal resource usage. Consequently, the crossbar structure is divided into three stages: input tagging, routing through the crossbar, and output indexing, as illustrated in \Cref{fig:fifoxbar}.

            It is important to note that routing \emph{any} set of values in this way is not always efficient. Consider, e.g., the extreme case where all tags in the input vector are identical. This would force all distribution modules to direct their inputs to the same output, causing back-pressure at the inputs of each stage. However, in practice we observe that the tags resulting from the check-node mapping to the \emph{$U,V$ tiles} are uniformly distributed, making this a non-issue in practice.
            
            There is nonetheless a slight penalty when flushing the crossbar interconnect, due to a small number of conflicts arising during processing. To mitigate this effect, all nodes have two input and two output AXI-Stream ports that can all be active simultaneously. FIFOs use a write-first dual port BRAM that can also bypass the input to the output register when required, while registers employ ping-pong buffers to allow for simultaneous and continuous I/O. Consequently, each distribution module routes four inputs to four outputs (corresponding to two FIFOs or ping-pong buffers), which greatly reduces the collision probability.

    \subsection{Decoder interconnect} 

        The final challenge lies in interconnecting the \emph{$D_X,D_Z$ unit} with the \emph{$U,V$ unit}. During the processing of $D_X$ and $D_Z$, a variable from $\bar{e}_X,\bar{e}_Z$ is considered ready once it is no longer be updated within the current iteration. At that point, it must be routed to the \emph{$U,V$ unit}. 
        Conversely, after being updated in the \emph{$U,V$ unit}, variables must be routed back to the \emph{$D_X,D_Z$ unit} for the next iteration.

             \begin{figure*}
            \centering
            \includegraphics[width=\linewidth]{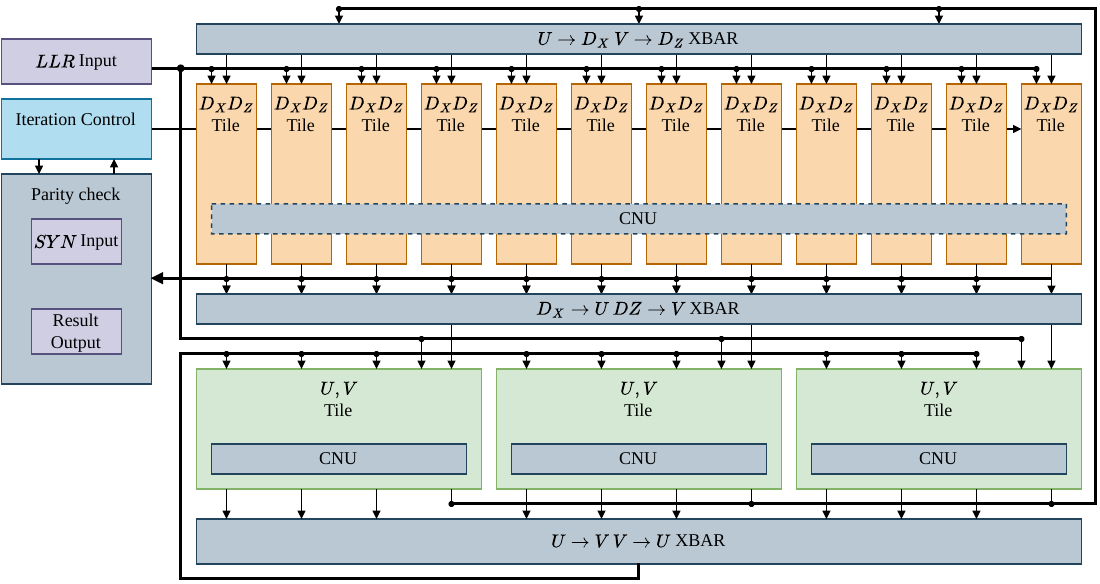}
            \caption{Architecture for the full decoder. Note the straightforward mapping between $D_XD_Z$ and $UV$ tiles.}
            \label{fig:fulldecoder}
        \end{figure*}
    
        Any updates related to $U,V$ can, in principle, be performed in any \emph{$U,V$ tile}. However, depending on the mapping, routing incurs a variable latency. Furthermore, not all checks in $U,V$ have the same degree, so an ideal mapping would assign checks of similar degree to the same \emph{$U,V$ tiles} in order to optimize resource usage. 
        In our implementation, a greedy strategy is used that maps the largest checks to the largest tiles, resulting in sufficiently low routing overhead that remains within the timing constraints imposed by the $D_X,D_Z$ unit. For the resulting \emph{$D_X,D_Z$ tiles} to \emph{$U,V$ tiles} routing, we use the same crossbar structure as that used for the $U\to V, V\to U$ feedback. 
        
        Routing back is equally simple by following the same process, with a different crossbar doing the inverse operation.
        
        Both routing steps are implemented using crossbars, as shown in \Cref{fig:fulldecoder}. Arrival ordering is not important at this stage, since the processing in the $D_X,D_Z$ and $U,V$ units is fully independent; data is therefore output in a round-robin fashion where necessary. Tagging is again used at the output of each stage to direct the messages through these modules. Queues are used where needed to avoid bottlenecks due to output readout availability. Synchronization of all data queues is necessary after each half-iteration. Since timing can be deterministically calculated, a counter suffices for this purpose, avoiding the need to monitor internal queues.

\section{Resource and timing estimation}\label{sec:restiming}

    In the following, we consider the decoding matrix $\bar{D}_{XYZ}$ as shown in \Cref{eq:GARI_matrix}, for the specific case of the $[[144,12,12]]$ gross code (parameters are as indicated in \Cref{tab:GARIsize}). While we provide numerical values for this code, the timing can be precisely estimated as a function of the number of checks. Resource usage, however, strongly depends on code-specific characteristics such as the check degrees in both $D_{X},D_{Z}$ and $U,V$, the number of variable nodes, the feasibility of optimal mappings to the architecture, etc. For this reason, a generalization of resource estimates is not provided. 

    \begin{table}[h]
        \centering
        \caption{Size of the different decoding sub-matrices for the $[[144,12,12]]$ gross code}
        \label{tab:GARIsize}
        \begin{tabular}{cccc}
             $D_X$ & $D_Z$ & $U$ & $V$ \\\hline\hline
             $792\times7920$ & $936\times8784$ & $7920\times 51048$ & $8784\times 51048$ \\ 
        \end{tabular}
    \end{table}

    The $D_{X}, D_{Z}$ processing core consists of 45 tiles. 
    This value is obtained from the best mapping found between variable nodes and tiles such that all checks can be executed and each variable node is assigned to a single tile. In addition, there are 18 \emph{$U,V$ tiles} with up to 1000 variable node each (to accommodate up to $18000 > 7920+8784$ nodes from $D_X$ and $D_Z$).
    The mapping is based on the degree of each variable node in the $U,V$ part, with higher degree nodes assigned to the lower-indexed \emph{$U,V$ tiles}, thereby optimizing resources. The degrees of the $U,V$ units are $(23, 17, 13, 11, 11, 11, 7, 7, 7, 7, 7, 7, 7, 7, 5, 5, 3, 3)$.  Each \emph{$U,V$ tile} receives up to 500 nodes from each $D_X$ and $D_Z$.

    The three crossbars are of sizes $45\to18$ (from $D_{X},D_{Z}\to U,V$), $18\to45$ (from $U,V\to D_{X},D_{Z}$) and $122\to122$ (from $U,V\to U,V$).

    The convergence check unit computes all $D_Z$ parity checks in parallel, aggregating a total of $936$ checks over $8784$ variables. 

    \subsection{Timing}
\begin{figure}
            \centering
            \includegraphics[width=\linewidth]{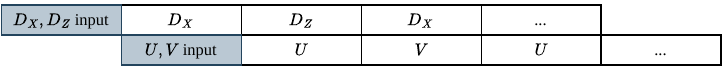}
            \caption{General timing diagram for the decoder.}
            \label{fig:timing_simpl}
        \end{figure}

        For timing, we consider the different processing steps and discuss how a pipelined architecture can overlap them to obtain the final estimate. A general timing  diagram is given in \Cref{fig:timing_simpl}.

        \begin{figure*}
            \centering
            \includegraphics[width=\linewidth]{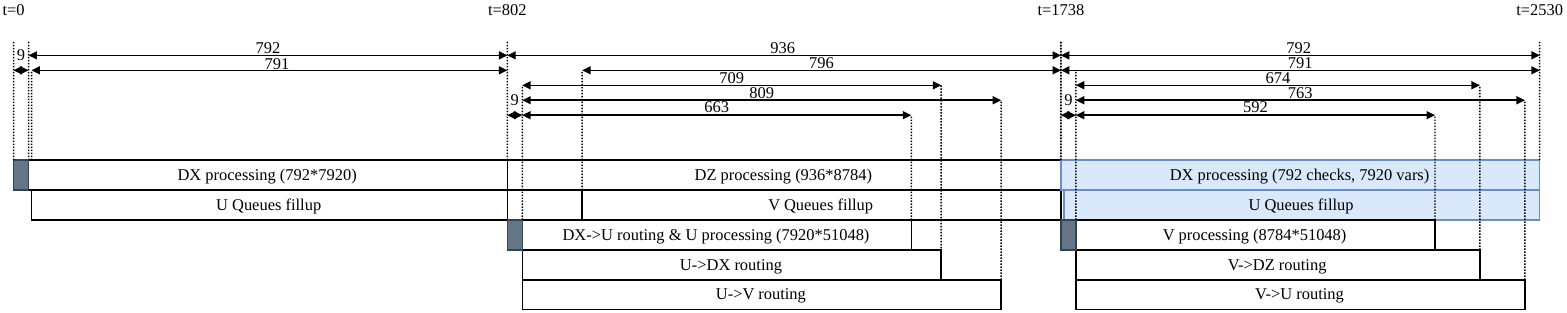}
            \caption{Timing for the $[[144,12,12]]$ gross code.}
            \label{fig:timing}
        \end{figure*}

        \subsection{Input}\label{subsub:input}
            The input to the decoder consists of two parts: the LLR values and the syndromes. We do not discuss how these values are obtained by the control unit of the quantum processor, and instead assume they are all available through registers and/or queues. Note that LLRs are not necessarily updated in each decoding round.

            LLR input throughput is limited by the number of memory banks available in each decoder. For the $D_X$ and $D_Z$, this number is given by the specific mapping of variables to \emph{$D_X,D_Z$ tiles}. In general, the mapping aims to minimize the number of tiles (that is, optimize resource usage) by placing independent variable nodes (i.e., that do not share checks) within the same tiles. However, this may increase LLRs loading time since each tile aggregates as many variable nodes as possible. An alternative approach is to increase the number of tiles, thereby reducing the per-tile load and improving parallel loading capability. In all cases, the minimum number of cycles required to fully load the LLRs is determined by the tile containing the largest number of variable nodes.

            Since processing in the \emph{$D_X,D_Z$ unit} proceeds sequentially, with the initial LLRs read only once, LLR loading can be overlapped with the computation of the previous decoding round. 
            This, together with the fact that input LLRs are not necessarily updated at every decoding round, allows the initialization overhead to be neglected.
            A similar argument applies to the initialization of the \emph{$U,V$ unit}, which can be overlapped with the first iteration of the \emph{$D_X,D_Z$ unit} to reduce input bandwidth requirements.

        \subsection{$D_X$ and $D_Z$ processing}

            Since the $D_X,D_Z$ decoding process proceeds sequentially over the checks, the timing is given by the number of checks plus a small overhead to fill the pipeline. 
            In our experiments, eight pipeline stages are sufficient to achieve clock frequencies above 300\,MHz. It is important to note that the separation between checks that share a variable must be greater than the number of pipeline stages to avoid using outdated (stale) values. In practice, for the $[[144,12,12]]$ gross code, this separation exceeds 10, and thus this constraint has not been observed to be limiting. 

        \subsection{$D_X\to U$ and $D_Z \to V$ message exchange}

            Messages from the \emph{$D_X, D_Z$ unit} are stored in the input queues to the $D_X\to U, D_Z\to V$ crossbar until the decoding iteration is completed. At that moment, message passing is initiated through the  crossbar. The crossbar has as many inputs as there are tiles in the \emph{$D_X, D_Z$ unit}, and as many outputs as there are tiles in the \emph{$U,V$ unit}. In general,  the former unit has more tiles than the latter, which results in a bottleneck at the output (alleviated by the fact that the output vectors often do not have all values enabled). 
            Experimentally, it has been found that the routing introduces an overhead of approximately $\approx 10-20\%$ (in number of cycles) relative to the number of messages received by each $U,V$ tile.  

        \subsection{$U$ and $V$ processing}
            Since the \emph{$U,V$ unit} is passive (i.e., it processes incoming values as they are available, with no control overhead), processing is naturally overlapped with the inputting of values, adding to the total timing the pipeline depth within the tiles. This is the same architecture as the one for \emph{$D_X,D_Z$ unit}, but in this case the pipeline depth is of 10 stages and the maximum check-node degree is different. 

        \subsection{$U \to V,D_X$ and $V\to U, D_Z$ message exchange}

            Messages coming from the \emph{$U,V$ unit} pass through two different crossbars: 
                \begin{itemize}
                    \item The feedback crossbar to the \emph{$U,V$ unit} is the most expensive, since each $U,V$ tile  produces a set of messages that may be routed to any other tile. Experimentally, this results in an overhead of around $20\%$ in cycles relative to the maximum number of checks on a single $U,V$ tile.  
                    \item The feedback to the \emph{$D_X,D_Z$ unit} has a lower overhead since the crossbar has more outputs than inputs. 
                    This results in only around a $10\%$ overhead relative to the maximum number of checks on a single $U,V$ tile. 
                \end{itemize} 

        \subsection{Total timing for the $U,V$ part}
            As we overlap the processing of $U$ and $D_Z$, and the processing of $V$ and $D_X$, the former two matrices must complete execution before the latter can proceed. To facilitate synchronization, data transfer from $D_X\to U$ and $D_Z\to V$ is enabled at the end of the $D_X$ and $D_Z$ processing stages, respectively. 

            This implies that, while processing $D_Z$, we must complete $D_X\to U$ routing, $U$ processing, and $U\to D_X,V$ routing (as shown in \Cref{fig:timing}). The same applies to the symmetric case of $D_X$. The total time  depends directly on the number of \emph{$U,V$ tiles} and the number of nodes processed by each tile. To ensure timing is tight enough, cycle-accurate simulations are performed to fine-tune these values, which in turn determine resource use.

        \subsection{Timing for the $[[144,12,12]]$ gross code}
            Timing in this architecture depends on several factors. First, the code determines the number of checks for each $D_X$ and $D_Z$. While the $U,V$ part can be parallelized since all checks within these matrices are independent, the benefits of parallelization may be partially offset by routing overhead in the crossbars. This overhead is also influenced by the variable mapping to both the $D_X,D_Z$ and $U,V$ \emph{tiles}.

            Considering all these factors, and since a hard limit is imposed by the checks in $D_X,D_Z$, the number of tiles and the mappings are chosen such that the timing of the remaining parts of the system is below this limit.

            Experimentally, for the [[144,12,12]] code after the GARI method from \cite{maan2025decodingcorrelatederrorsquantum}, 
            a grouping has been found with a maximum size of 345 variable nodes 
            per tile in $D_X$ and 286 in $D_Z$, resulting in $345+286=631$ cycles for loading. This result was obtained using a guided heuristic search over the solution space, terminated once a sufficiently good solution was found. Finding the optimal grouping is NP-hard and out of the scope of this paper. Furthermore, as explained in \Cref{subsub:input}, input loading can overlap with other processing stages, making this overhead non-critical for overall performance.

            For the \emph{$U,V$ tiles}, 500 variable nodes 
            are mapped to each tile (keeping the count below 512 allows for some resource optimization in terms of BRAM usage). In total, 18 tiles are required for the complete mapping of $D_Z$, which is the largest of the two matrices with 8784 variable nodes.

            Simulations have been carried out to fully estimate the timing of each part of the system, yielding the results shown in \Cref{fig:timing}. Given the 500 nodes per \emph{$U,V$ tile}, the routing overhead remains well within the constraints imposed by the $D_X,D_Z$ processing. 

            Processing all checks in $D_Z$ and $D_X$ takes $936$ and $792$ cycles, respectively, yielding a total of $936+792=1728$ cycles per iteration. For the first iteration, the pipeline fill latency of $10$ cycles must also be accounted for. Thus, for $i$ iterations, the total cycle count is:
            \begin{equation}
                10 + 1728\cdot i
            \end{equation}
            We have implemented the architecture on an AMD VU29P FPGA with a clock period of $3.647$\,ns, corresponding to a clock frequency of $\approx274$\,MHz. This yields a total decoding latency of:
            %
            \begin{equation}
                (1728\cdot i +10) \cdot 3.647\,\text{ns} = i\cdot 6302\,\text{ns} + 37\,\text{ns}
            \end{equation}
            %
            Assuming a physical error rate of $0.001$ for the $[[144,12,12]]$ code, the average number of decoding iterations for a single decoder that operates on a window of $d=12$ consecutive syndrome measurements \cite{maan2025decodingcorrelatederrorsquantum} is $2.28$, yielding an average latency of $2.28 \cdot 6302\,\text{ns}+37\,\text{ns}=14.4\,\mu\text{s}$, and $1.13$ iterations for an ensemble of 24 decoders, yielding an average latency of $1.13 \cdot 6302\,\text{ns}+37\,\text{ns}=7.16\,\mu\text{s}$. The input LLRs are quantized to 6 bits, check-node messages to 8 bits, and variable-node messages to 10 bits. Using these quantization parameters, we closely approach the accuracy of the floating-point implementation. 

%

        \subsection{Resources for the $[[144,12,12]]$ gross code}



\begin{figure}
                \centering
                \includegraphics[width=\linewidth]{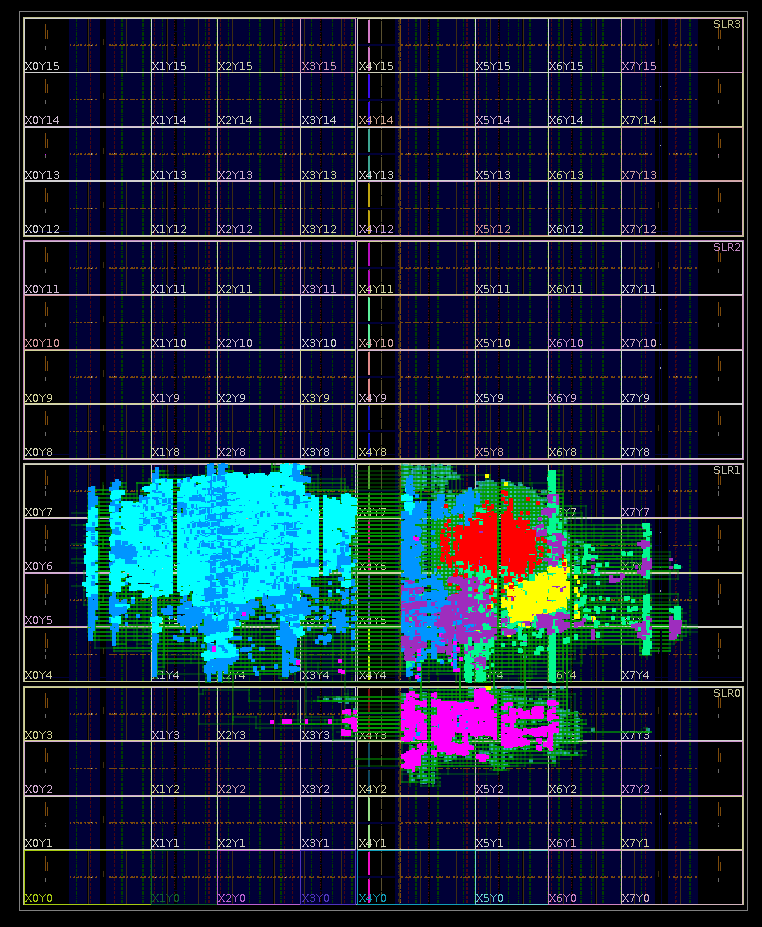}
                \caption{Placement of a single core of the proposed decoder for correlated errors within the FPGA. $D_{X}, D_{Z}$ core (yellow), Vnode memory (green), Parity check (red), $D_{X},D_{Z}\to U,V$ (pink), $U,V\to D_{X},D_{Z}$ (purple), $U,V$ unit (dark blue), $U,V\to V,U$ (light blue).}
                \label{fig:utilization}
            \end{figure}

            The high-level decoder architecture is illustrated in \Cref{fig:fulldecoder}, and the resource usage of the different modules therein is given in \Cref{tab:resourcesestimation}. The placement of the modules across the FPGA is also shown in \Cref{fig:utilization}.

            \begin{table}[h]
                \centering
                \caption{Resource usage of a single decoder for correlated errors for the $[[144,12,12]]$ gross code in an AMD's VU29P FPGA}
                \label{tab:resourcesestimation}
                \begin{tabular}{c|c|c|c}
                    Module & LUTs & REGs & BRAMs \\\hline\hline
                    Full core               & 122393 & 111697 & 704 \\\hline
                    $U,V\to D_{XZ}$ XBAR    & 11650 & 8296  & 31.5  \\
                    $U,V\to V,U$ XBAR       & 55645 & 41059 & 125   \\
                    $D_{XZ}\to U,V$ XBAR    & 8858  & 6844  & 25    \\
                    $D_{XZ}$ core (degree 45) & 11451  & 7961  & 70.5     \\
                    $D_{X},D_{Z}$ memory (each)       & 2241  & 770   & 80.5  \\
                    \emph{$U,V$ tile} (degree 23) & 3767 & 1186 & 44.5 \\
                    \emph{$U,V$ tile} (degree 17) & 2587 & 882 & 32 \\
                    \emph{$U,V$ tile} (degree 11) & 1690 & 573 & 20 \\
                    \emph{$U,V$ tile} (degree  5) &  907 & 386 & 12 \\
                    Parity check                  & 5789 & 936 & 0     \\\hline
                    Others (interconnect logic)   & rest & rest & rest \\
                \end{tabular}
            \end{table}

            For the \emph{$D_{X},D_{Z}$} and \emph{$U,V$ units} we report both the processing cores and the associated memories (used for storing messages and tagging them with their corresponding addresses). 
            Note that only a subset of the \emph{$U,V$ tiles} 
            is reported for simplicity and clarity. For the crossbars, we report only the logic used to redistribute values, and not the tagging logic, which is performed within the tiles. Finally, for the parity check unit we report the full parallel computation of $D_Z$ parity checks. Other smaller modules are omitted for simplicity.

            The full core, implemented on an AMD VU29P FPGA, occupies  7.5\% of the total LUTs, 3.5\% of the total REGs, and 26\% of the total BRAMs. Alternatively, we have mapped part of the BRAMs to URAMs. In this case, the designs achieves  21\% RAM usage and 5\% URAM usage.
            
\begin{figure}
                \centering
                \includegraphics[width=\linewidth]{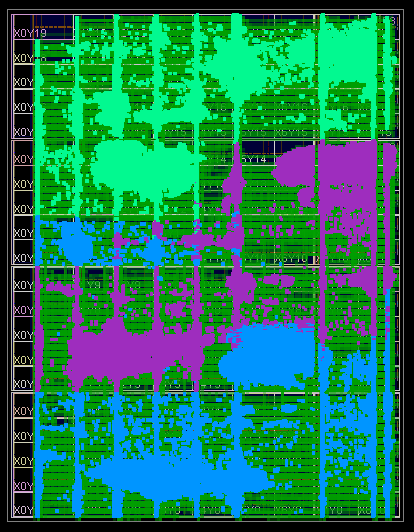}
                \caption{Placement of a three-decoder ensemble for correlated errors within the FPGA. Each core is highlighted in a different color (light green, purple, and blue). The dark green lines show the routing between the different processing units. } 
                \label{fig:utilization_ensemble}
            \end{figure}

           As can be seen, there is sufficient space to implement multiple cores on the same device, which enables reductions in power consumption and classical resources, including inter-board communication. In this scenario, the limiting factor is the available BRAMs. To maintain the same operating frequency as a single core, and ensure compliance with real-time constraints, three instances were implemented on the same device. The resulting resource usage is reported in \Cref{tab:resourcesestimationensemble}, and the corresponding layout is shown in \Cref{fig:utilization_ensemble}.

           \begin{table}[h]
                \centering
                \caption{Resource usage of a three-decoder ensemble for the $[[144,12,12]]$ gross code in an AMD VCU19P FPGA}
                \label{tab:resourcesestimationensemble}
                \begin{tabular}{c|c|c|c}
                    Module & LUTs & REGs & BRAMs \\\hline\hline
                    Ensemble of size 3              & 	381698 &	365765 & 2121 \\\hline
                \end{tabular}    
            \end{table}

This means that, for the complete ensemble of  24 decoders used in \cite{maan2025decodingcorrelatederrorsquantum}, only eight VCU19P devices are required, compared to the 48 devices needed in the parallel architecture proposed in the same work, while still satisfying the timing constraints. 

Furthermore, the proposed architecture can accommodate three decoders for correlated errors for the [[144,12,12]] code, whereas \cite{maurer2025realtimedecodinggrosscode} reports only a single instance of an uncorrelated decoder. This indicates that the proposed design can achieve higher decoding capacity on the same device by exploiting the GARI method.

\section{Conclusions}
This paper presented a hardware architecture for decoding correlated errors in quantum LDPC codes by exploiting the structure of the detector error model after the GARI transformation. The proposed design balances parallelism while preserving scalability, achieving low latency with moderate area and power consumption. This enables the integration of multiple decoding cores within a single device, facilitating ensemble-based decoding to improve the logical error rate.

For the [[144,12,12]] bivariate bicycle code, three instances of the proposed core were implemented on a VCU19P FPGA to process the complete graph with $X$, $Y$, and $Z$ errors. The implementation achieves an average latency of 7.15\,$\mu$s over 12 rounds ($7.15\mu$s$/12=596$ns per round), utilizing eight FPGAs, while occupying 9\% of LUTs and 4\% of registers, and 90\% of BRAMs. 
These results demonstrate that decoding of correlated-error graphs and multi-core ensemble configurations can be efficiently realized by implementing several cores within a single device, 
improving upon prior hardware approaches with a sixfold reduction in hardware resources.

Future work will address the integration of full ensemble decoding and real-time noise emulation to evaluate accuracy at low logical error rates. In addition, the exploration of heterogeneous acceleration in conjunction with FPGA platforms or ASIC implementations will be considered to further scale the proposed solution while meeting latency and power requirements, with the goal of integrating the entire ensemble within a single chip.



%

\section*{Acknowledgment}

D.B.G and F.G.H. acknowledge support from the project PID2023-147059OB-I00 funded by MCIU/ AEI/ 10.13039/501100011033/ FEDER.  UE. F.G.H. work was also partially funded by a grant from Google Quantum AI.

\ifCLASSOPTIONcaptionsoff
  \newpage
\fi



%

\bibliographystyle{IEEEtran}

\bibliography{{bibtex/bib/IEEEexample}}

%








\end{document}